\documentclass[aps,prl,reprint]{revtex4-1}
\usepackage{amsmath,amsfonts,amssymb}
\usepackage{graphicx,bm}
\usepackage[breaklinks,colorlinks,urlcolor=blue,citecolor=blue,linkcolor=blue]{hyperref}
\begin{document}
\title{Quantum emulation of topological magneto-optical effects using ultracold atoms}
\author{Zhen Zheng}\email{zhenzhen.dr@outlook.com}
\author{Z. D. Wang}\email{zwang@hku.hk}
\affiliation{Guangdong-Hong Kong Joint Laboratory of Quantum Matter, Department of Physics, and HKU-UCAS Joint Institute for Theoretical and Computational Physics at Hong Kong, The University of Hong Kong, Pokfulam Road, Hong Kong, China}
%----------------------------------------------------------------------------------------
\begin{abstract}

Magneto-optical effect is a fundamental but broad concept in magnetic mediums.
Here we propose a scheme for its quantum emulation using ultracold atoms.
By representing the light-medium interaction in the quantum-emulation manner,
the artificial magneto-optical effect emerges under an entirely different mechanism from the conventional picture.
The underlying polarization state extracted in the synthetic dimension displays a different response to various experimental setups.
Notably, the magneto-optical rotation is related to the bulk topology in synthetic dimensions,
and thus provides an unambiguous evidence for the desired topological magneto-optical effect, which has not been developed hitherto in ultracold atoms.
This scheme is simple and feasible, and can be realized by current experimental techniques.
The implementation of the scheme is able to offer an intriguing platform for exploring topological magneto-optical effects and associated physics.

\end{abstract}
\maketitle
%----------------------------------------------------------------------------------------

{\noindent\bf \MakeUppercase{Introduction}}

Magneto-optical (MO) effect is a broad concept that the polarization state of the light field is altered in response to the magnetization of mediums \cite{Ebert1996rpp}.
In condensed-matter systems, the physical origin of the MO effect is usually ascribed to the transverse conductivity of magnetic mediums \cite{Pershan1967jap}.
It is introduced by the intrinsic magnetization that originates from the presence of both the band exchange splitting and spin-orbit coupling \cite{Ebert1996rpp},
and hybridizes the two polarization states of the light field during the propagation in the medium.
Notably, the MO effect has recently attracted intensive interests with wide applications on the magnetization, e.g. the detection and manipulation to the magnetic order \cite{Kirilyuk2010rmp},
and the visualization of magnetic domains \cite{McCord2015jpd,Higo2018nphon}.

However, such a fundamental concept, particularly the topology-related one, has not been addressed in ultracold atoms.
This is because the light field usually stimulates atoms to excited states.
During the atomic level transition, the polarization of the light field normally remains unchanged
and hence the MO effect does not occur.
Furthermore in ultracold atomic gases, because of neutral charges, the kinetic motion of atoms is independent of the classic magnetization.
Therefore it is also frustrated to support the MO effect via the intrinsic magnetization of the atomic systems.
On the other hand, neutrally charged atomic gases have triggered the quantum emulation and engineering of artificial magnetic fields \cite{Lewenstein2007aip,Safronova2018rmp,Georgescu2014rmp},
and provided various applications on the synthetic topological materials and phases \cite{Dalibard2011rmp,Goldman2014rpp,Lin2016jpb,Zhang2018advphys,Cooper2019rmp}.

Here we present a feasible and systematic proposal for engineering a topological MO effect (TMOE) using ultracold atoms.
Different from a traditional picture,
we propose an entirely new mechanism for rotating the light polarization in a quantum-emulation manner.
This supports the analogy to the natural MO effect,
and manifests its topological features using artificially controllable techniques of ultracold atoms.
The main results of this work are as follows.
(i) The hybridization of polarization states traditionally induced by the transverse conductivity is now realized by the effective atomic current in a synthetic dimension, such that
the distinct phenomena in ultracold atoms and condensed-matter systems may stem from the same concept,
which also provides a new perspective for understanding the interplay between atoms and optical fields.
(ii) The polarization rotation of the emergent MO effect is intrinsically related to the topological invariant defined in the synthetic dimensions.
This origin of TMOE in ultracold atoms differs from TMOE in condensed-matter systems \cite{Tse2010prl,Tse2011prb,Tse2010prb,Feng2020natcommun,Okada2016natcommun,Wu2016sci,Dziom2017natcommun},
in which the transverse conductivity is related to the topological invariant instead.
(iii) The observed polarization rotation can exhibit different behaviors between the topological and trivial MO effects, and thus it may provide a dual application for detecting the topological invariant.
(iv) Since the magnitude of the TMOE rotation is artificially controllable, this quantum emulation can avoid the practical problems encountered in the natural MO effect.
For example due to the photon absorption of the medium, the traditional measurement is usually performed in ultra-thin films and the magnitude of the accumulated MO rotation is thereby limited for the natural MO effect.
(v) The proposal can be realized with currently available experimental techniques of ultracold atoms.

{~\\\noindent\bf \MakeUppercase{Results}}

{\noindent\bf Model}

In a ultracold atomic gas under the macroscopic motion,
the dynamic evolution of the atomic number density is described by its continuity equation \cite{Pethick2008book,Zheng2021njp}, which inspires us with a reminiscent of the light propagation.
In particular, the atomic cloud can be used to mimic the light field
and its center-of-mass (COM) velocity serves as the light speed.
In ultracold atoms,
we choose $N_s$-fold atomic internal states as the pseudo-spin states,
which can be coupled via optical fields.
If the polarization of the emulated light could be represented by the pseudo-spin degrees of freedom,
it may provide a way for exhibiting the MO effect
by regarding the pumped area as the ``medium''.
At this stage, the mechanism of the artificial MO effect is rooted in the interplay between the atoms and optical fields.

\begin{figure}[t]
	\centering
	\includegraphics[width=0.48\textwidth]{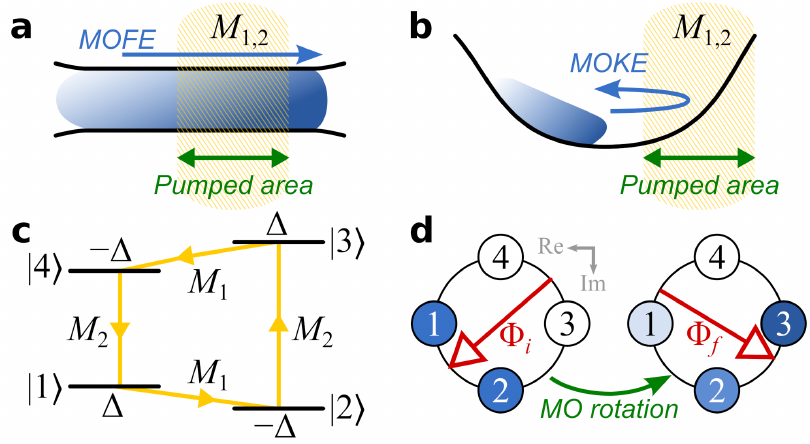}
	\caption{
		{\bf Illustration of the model setups.}
		The atomic current (emulated light) is prepared by ({\bf a}) the transport in a channel or ({\bf b}) the mechanical oscillation within a harmonic trapped potential.
		When entering the pumped area (yellow hatched area, the emulated medium), atomic pseudo-spin states are coupled by optical fields $M_{1,2}$.
		({\bf c}) Transitions between pseudo-spin states driven alternately by the optical fields $M_1$ and $M_2$.
		The pseudo-spin states are labeled by $|j\rangle$ ($j=1,2,3,4$) that simultaneously characterizes the position in the synthetic dimension.
		({\bf d}) The polarized angle $\Phi$ (red arrows) for the emulated light is defined by the COM coordinate in the synthetic ring lattice (constructed by the pseudo-spin states labeled by numbers).
		After the emulated light passes through or is reflected by the emulated medium, MOFE or MOKE occurs if the initial polarized angle $\Phi_i$ is rotated to the final one $\Phi_f$.
		The MO rotation is detected by the occupation in each pseudo-spin state (characterized by the blue color).
		We use the subscripts $i$ and $f$ as the denotations of the initial and final cases.
	}
	\label{fig:setup}
\end{figure}

In the natural MO effect,
the polarization of both the forward and reflected light fields
can process the rotation under the light-medium interaction.
The MO effect is thus specified in two representative ones,
respectively known as the MO Faraday effect (MOFE) and Kerr effect (MOKE) \cite{Argyres1955pr}.
In order to propose a systematic and complete scheme for emulating the MO effects,
we consider the following two protocols shown in Fig.\ref{fig:setup}{\bf a} and {\bf b}.
(i) The atoms are prepared in a channel containing the pumped area between two number-imbalanced reservoirs (e.g. \cite{Hausler2021prx}).
This protocol supports the steady atomic transport and is thus used for mimicking MOFE when the current passes through the pumped area.
(ii) The atoms are placed in regions deviated from the center of a harmonic trap potential.
The resultant mechanical oscillation makes atoms enter and leave the pumped area in periods,
which can be used for mimicking MOKE.

Extracting the definition of the polarization states by the pseudo-spins is crucial for signaling the potential MO effects.
To capture a clear picture, we focus on a simple case of $N_s=4$.
In the atoms, each pseudo-spin state is prepared to be coupled to the other two via optical fields,
forming an enclosed ring transition shown in Fig.\ref{fig:setup}{\bf c}.
If we represent the pseudo-spin states as the spatial coordinates,
the atoms can be regarded as being loaded in a ring lattice of the synthetic dimension.
In the polar coordinate frame,
the azimuth angle $\phi$ is discretized by $N_s$: $\phi_j=(j-1)\times2\pi/N_s$,
where $j=1,\cdots,N_s$ stands for the site(pseudo-spin) index.
We can use the complex variable
\begin{equation}
 	\hat{X}_j=e^{i\phi_j} \label{eq:coordinates}
\end{equation}
to denote the coordinate on the ring lattice.
This denotation has the advantage that it preserves the intrinsic periodicity of the ring geometry,
i.e. $\hat{X}_{j+N_s}=\hat{X}_j$.
The corresponding COM coordinate is thus obtained by
\begin{equation}
	\hat{X}_{\rm cm} = \sum_{j} \hat{X}_j \hat{n}_j = {\rm Tr}(\hat{X} \hat{n}) \,,
	\label{eq:com-coordinate}
\end{equation}
where $\hat{n}$ denotes the density matrix,
and the element of the coordinate matrix is defined as $[\hat{X}]_{ij} = \hat{X}_j\delta_{ij}$.

From the complex coordinate (\ref{eq:coordinates}), the two polarization states can be separately figured out by the real and imaginary parts of $\hat{X}_{\rm cm}$.
In this picture, the definition of the polarized angle $\Phi$ is extracted from the argument angle of $\hat{X}_{\rm cm}$ as follows,
\begin{equation}
	{\rm Arg}[ \hat{X}_{\rm cm} ] = \Phi \pm n\pi \,,\quad
	n=0,1,2,\cdots \label{eq:polarized-angle}
\end{equation}
In Eq.(\ref{eq:polarized-angle}), we have assumed that $0\leqslant\Phi<\pi$ and the polarization extracted from $\pm\hat{X}_{\rm cm}$ is identical,
which are inherited from the original properties of the natural polarized angle.
Under the viewpoint of the ring lattice,
the coupling between different pseudo-spin states can be regarded as the nearest-neighbor (NN) hopping.
As mentioned before, the microscopic origin of the natural MO effect is rooted in the transverse conductivity that hybridizes and rotates the polarization states.
In our scheme,
the NN hopping has the possibility to trigger the atomic current between NN sites,
and the resultant atomic macroscopic motion in the synthetic dimension
may lead to the MO rotation with changed $\Phi$,
as illustrated in Fig.\ref{fig:setup}{\bf d}.

However, for a system whose Hamiltonian is constructed solely by the homogeneous NN hopping, we observe no MO effect.
This is because the Hamiltonian is invariant
under the inversion operation $\hat{X} \rightarrow 2\hat{X}_{\rm cm}-\hat{X}$ with respect to $\hat{X}_{\rm cm}$.
During the evolution, the atomic density profile extends and is divided into two currents with identical number densities but in opposite directions on the ring lattice.
As a result, the MO rotation is conserved to be zero.

Generally, since the coupling between different pseudo-spin states is introduced via optical fields,
it offers a feasible tool for artificially designing the NN hopping beyond the homogeneous one.
In particular, one may prepare a staggered pattern for the coupling strength (as shown in Fig.\ref{fig:setup}{\bf c}),
and make the kinetic energy negligible in comparison to the coupling strength.
In the synthetic dimension,
a model Hamiltonian of the system reads,
\begin{align}
	H &= \sum_{j=1}^{N_s/2}(M_1 \psi^\dag_{2j} \psi_{2j-1}
	+ M_2 \psi^\dag_{2j+1} \psi_{2j} + H.c.) \notag\\
	&+\Delta (\psi_{2j-1}^\dag \psi_{2j-1}-\psi^\dag_{2j} \psi_{2j})
	\,, \label{eq:model-hamit}
\end{align}
which is also known as the Rice-Mele model \cite{Rice1982prl} but in a periodic boundary condition and a short lattice size.
Here $\Delta$ denotes the energy offset between atoms in adjacent sites.
$M_{1,2}$ denotes the coupling strength under the staggered pattern.
$\psi_j$ and $\psi_j^\dag$ stand for the annihilation and creation operators of the atom occupied on the coordinate $\hat{X}_j$.
We associate $\Delta$ and $M_{1,2}$ with the following cyclic pump,
\begin{equation}
	\begin{cases}
		M_1(t) = M_0-\alpha\sin(\nu t) \\
		M_2(t) = M_0+\alpha\sin(\nu t) \\
		\Delta(t) = \Delta_0 + \alpha'\cos(\nu't + \varphi_0)
	\end{cases}
	\label{eq:params-pump}
\end{equation}
where $\alpha^{(\prime)}$ and $\nu^{(\prime)}$ are the amplitude and frequency for the pump, respectively.
$\varphi_0$ is the relative phase.
Hereafter we choose $\alpha$ as the energy unit.
The pump period of $M_{1,2}$ is given by $T=2\pi/\nu$,
and we choose it as the time unit.
Specifically for the steady transport in Fig.\ref{fig:setup}{\bf a},
the total pumped time $t_{\rm tot}=L/v_0$ can be evaluated by the width $L$ of the pumped area and the current velocity $v_0$.

We first investigate the case with $\nu=\nu'$,
and introduce a dimensionless parameter $\varphi\equiv\nu t$.
Such a cyclic pump is also known as the Thouless quantum pump \cite{Thouless1983prb,Niu1984jpa}, and has been theoretically \cite{Romero-Isart2007pra,Qian2011pra,Wang2013prl-1,Wang2013prl-2,Zeng2015prl,Taddia2017prl,Ke2017pra,Marra2020prr} and experimentally \cite{Lohse2016nphys,Nakajima2016nphys} studied using ultracold atoms but in real spaces.
We can see that Hamiltonian (\ref{eq:model-hamit}) is invariant under the operation $\varphi \rightarrow \varphi+2\pi$.
It indicates that $\varphi$ can be recognized as the pseudo-momentum in an auxiliary dimension of the reciprocal lattice.
Consequently, the one-dimensional system is mapped into the two-dimensional one,
in which the Brillouin zone is defined in the $k_{X}$-$\varphi$ plane instead.
This paves the way for defining the Berry phase for Hamiltonian (\ref{eq:model-hamit}) \cite{Atala2013nphys},
\begin{equation}
	\gamma_{\pm} = \frac{1}{2\pi} \int \Omega_{\pm}(k_{X},\varphi) d k_{X}d\varphi \,,
\end{equation}
where $\pm$ denotes the band index,
$\Omega_{\pm}$ is the Berry curvature:
\begin{equation}
	\Omega_{\pm}(k_{X},\varphi) = i
	\frac{\langle\pm| \nabla_{\varphi} H |\mp\rangle\langle\mp| \nabla_{k_{X}}H |\pm\rangle}{(E_{\pm}-E_{\mp})^2}
	- (k_{X} \leftrightarrow \varphi)
\end{equation}
and $E_{\pm}$ is the energy of $|\pm\rangle$.

\begin{figure}[t]
	\centering
	\includegraphics[width=0.48\textwidth]{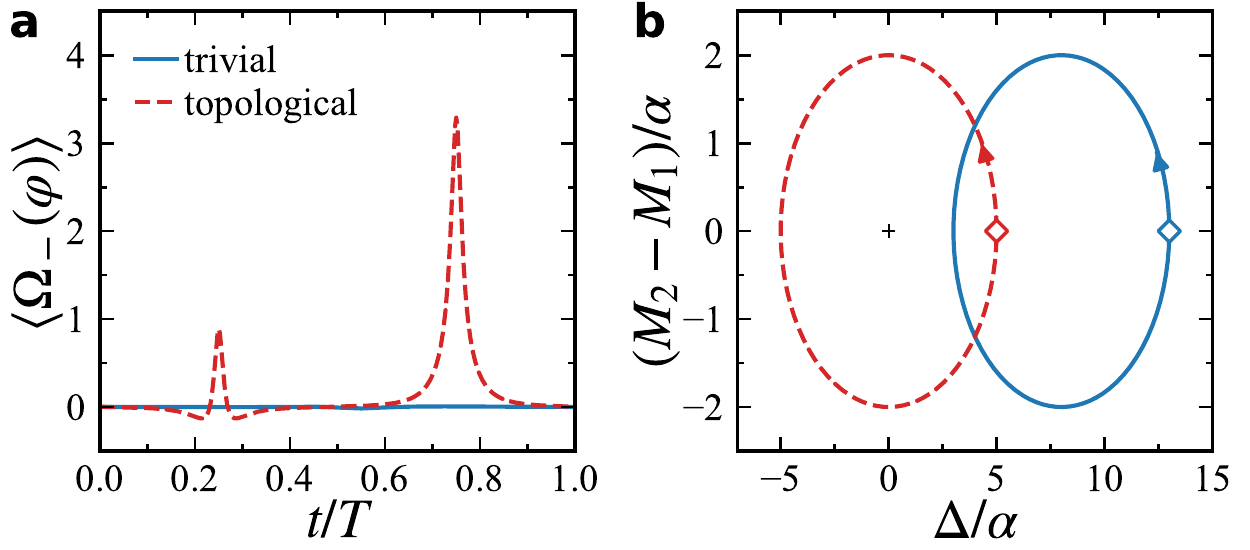}
	\caption{
		{\bf Dynamic evolution in one period of pump cycling.}
		({\bf a}) Evolution of the averaged Berry curvature $\langle \Omega_-(\varphi) \rangle$.
		The blue-solid and red-dashed lines show the cases of the trivial pump with $\Delta_0=8\alpha$
		and the topological pump with $\Delta_0=0$, respectively.
		The pump starts at $t_i=0$ and ends at $t_f=T$.
		We set $M_0=-0.2\alpha$, $\alpha'=5\alpha$, and $\varphi_0=0$.
		({\bf b}) The corresponding trajectories of the evolution in the parameterized plane between $\Delta(t)$ and $M_2(t)-M_1(t)$.
		The diamonds mark the positions before the pump,
		and the arrows show the directions of the trajectories.
		The plus symbol marks the singular point where the Berry phase is ill defined.
		Only the trajectory of the topological pump encloses the singular point.
	}
	\label{fig:berry-curvature}
\end{figure}

The nontrivial Berry phase reveals an alternative way for generating the atomic current.
Notice that in Eq.(\ref{eq:model-hamit}), the effective chemical potential is zero.
Only the lower band $|-\rangle$ is fully occupied.
Therefore, the triggered topological current can be evaluated by the Berry phase $\gamma_-$.
In Fig.\ref{fig:berry-curvature}, we display the evolution of averaged Berry curvature $\langle \Omega_-(\varphi) \rangle \equiv \int \Omega_-(k_X,\varphi)d k_X$ in one period of pump cycling.
There are two types of the cyclic pump modes, classified by $\gamma_-$.
For the trivial pump, the singular point with an ill-defined Berry phase \cite{gapless-footnote} is out of the evolution trajectory, as shown in the parameterized plane of Fig.\ref{fig:berry-curvature}{\bf b}.
$\langle \Omega_-(\varphi) \rangle$ insensitively oscillates in the vicinity of zero and our calculation gives $\gamma_-=0$.
It indicates that COM of the atomic cloud will preserve its initial position after one period of pump cycling.
By contrast, for the topological pump whose trajectory counterclockwise encloses the singular point,
we can see $\langle \Omega_-(\varphi) \rangle>0$ in certain regimes of Fig.\ref{fig:berry-curvature}{\bf a}, yielding $\gamma_-$ is nonzero
(in fact our calculation gives the quantized result $\gamma_-=1$).
It indicates that COM will move forward in the synthetic ring lattice.
Therefore, by preparing the pump mode,
the possible MO rotation can be triggered by the current in the topological pump,
and is thereby related to $\gamma_-$.
In this sense, a TMOE will be achieved.

{~\\\noindent\bf Topological MO rotation}

\begin{figure}[t]
	\centering
	\includegraphics[width=0.48\textwidth]{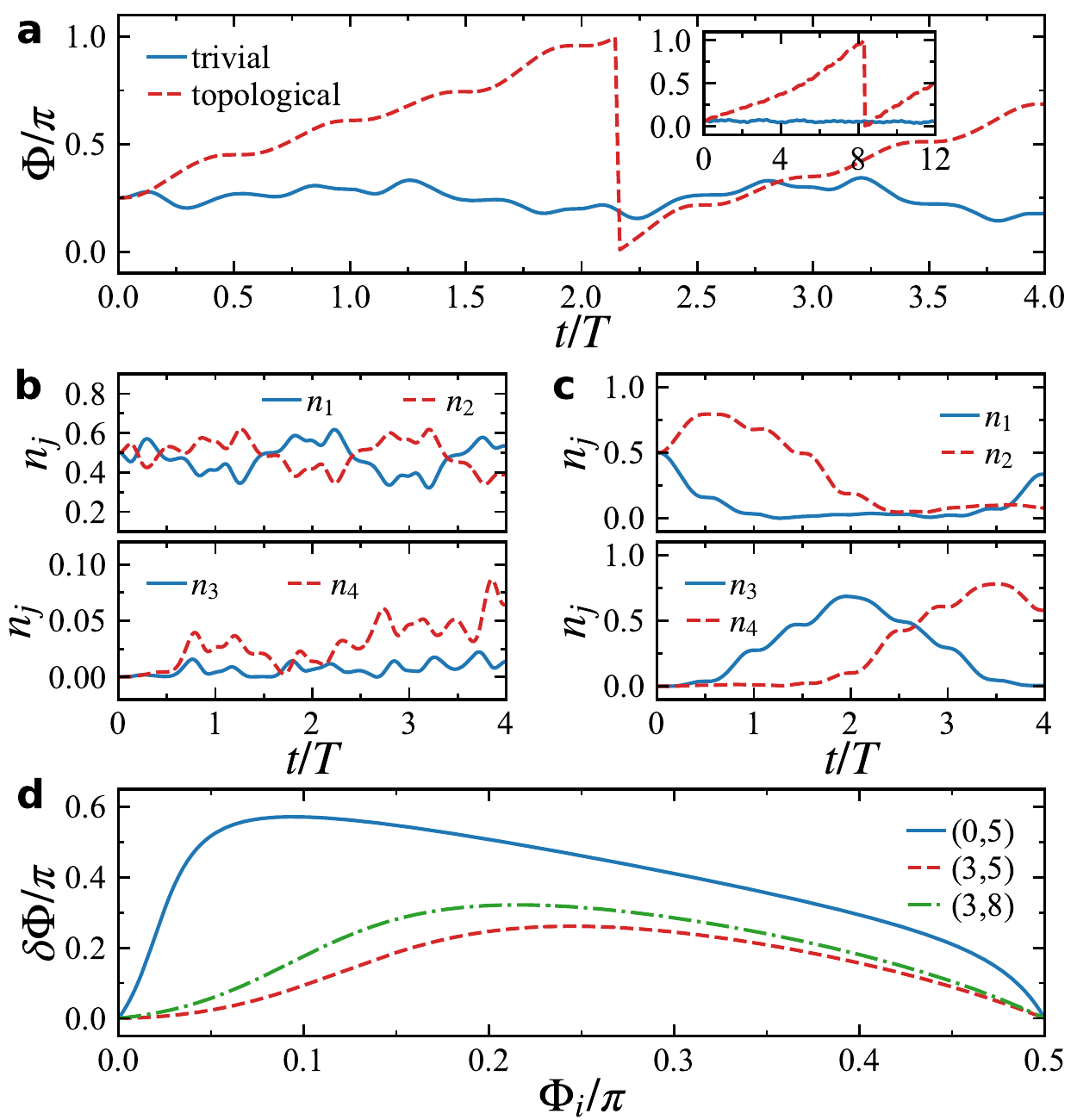}
	\caption{
		{\bf Numerical results of the MO rotation.}
		({\bf a}) Evolution of the polarized angle $\Phi$ for the trivial (blue-solid lines) and topological pump (red-dashed lines).
		We set the initial polarized angle $\Phi_i=\pi/4$ at $t_i=0$.
		Other parameters are the same with Fig.\ref{fig:berry-curvature}.
		The inner panel of ({\bf a}) shows the case of $N_s=16$.
		({\bf b})-({\bf c}) The evolution of the atomic densities in the ({\bf b}) trivial and ({\bf c}) topological pump.
		$n_j$ stands for the density occupied in $|j\rangle$.
		({\bf d}) The topological MO rotation $\delta\Phi=\Phi_f-\Phi_i$ as a function of the initial polarized angle $\Phi_i$ for various parameter sets ($\Delta_0$,$\alpha'$).
		The evolution processes four periods of pump cycling (i.e. $t_f-t_i=4T$).
	}
	\label{fig:evol-equal-freq}
\end{figure}

The atoms start the dynamic evolution after the atomic current (emulated light) enters the pumped area (emulated medium).
The time-dependent evolution of $\Phi$ can be analyzed by numerical simulations (see ``METHODS").
The results are shown in Fig.\ref{fig:evol-equal-freq}{\bf a}.
For the trivial pump,
the atoms preserve the dominant occupation in initial pseudo-spin states even after entering the emulated medium.
They exhibit a Rabi oscillation between the two states as shown in Fig.\ref{fig:evol-equal-freq}{\bf b}.
Therefore,
$X_{\rm cm}$ insensitively oscillates centered at the initial position,
and $\Phi$ varies around a narrow vicinity of $\Phi_i$.
For the topological pump, noticing that $\Phi=0$ and $\pi$ are equivalent,
$\Phi$ in fact increases continuously and is consistent with the positive $\gamma_-$.
This is also shown by Fig.\ref{fig:evol-equal-freq}{\bf c},
in which the atomic occupation processes the counterclockwise current in the ring lattice of Fig.\ref{fig:setup}.
Such a distinct result of TMOE renders its observation more tractable.
As the final polarization is determined by the total pumped time $t_{\rm tot}$,
the MO rotation changes quantitatively under various setups when the current leaves the pumped area.
However, the observable rotation only occurs in TMOE,
which is qualitatively distinguished from the trivial case.

The topological MO rotation also depends on the relative phase $\varphi_0$ in Eq.(\ref{eq:params-pump}).
For example, if we replace the set of $\varphi_0=0$ in Fig.\ref{fig:berry-curvature} by $\varphi_0=\pi$,
the evolution trajectory will have the same shape as in Fig.\ref{fig:berry-curvature}{\bf b},
but its direction changes to be clockwise, yielding a negative $\gamma_-$.
At this time, COM of the atomic cloud moves along the clockwise direction of the synthetic ring lattice, leading to the reversal MO rotation.

In Fig.\ref{fig:evol-equal-freq}{\bf d},
we show the topological MO rotation $\delta\Phi=\Phi_f-\Phi_i$ with respect to the initial polarized angle $\Phi_i$ after four periods of pump cycling.
We can see that $\delta\Phi$ continuously depends on the incoming polarized angle $\Phi_i$,
which is a unique feature of the artificial TMOE compared with the natural MO effect.
Furthermore, $\delta\Phi$ is also affected by the parameters $\Delta_0$ and $\alpha'$
and thereby is controllable via experimental methods.
Since the rotation occurs in the synthetic dimension,
its detection is universal and robust under various spatial distributions of atoms.

The previous results are obtained by the synthetic lattice of four sites.
In the inner panel of Fig.\ref{fig:evol-equal-freq}{\bf a}, we study the influence of the lattice size $N_s$ on the MO effect.
It is qualitatively seen that the measurement of topological MO rotation is robust against the finite-size effect.
Since the synthetic lattice hosts the ring geometry,
there is no boundary effect on the measurement, which is an advantage for observing TMOE.
However, with the increase of $N_s$, it requires more periods of pump cycling to observe the same rotation.
This is because from the definition of the complex coordinate (\ref{eq:coordinates}),
it is known that the $2\pi$ range of $\Phi$ is discretized by $N_s$.
Hence the periodicity, during which the polarization returns to the initial value, is proportional to $N_s$.
This can be approximately extracted from Fig.\ref{fig:evol-equal-freq}{\bf a}.
It reveals that the larger lattice size can perform a more precise measurement of the topological MO rotation.

\begin{figure}[t]
	\centering
	\includegraphics[width=0.48\textwidth]{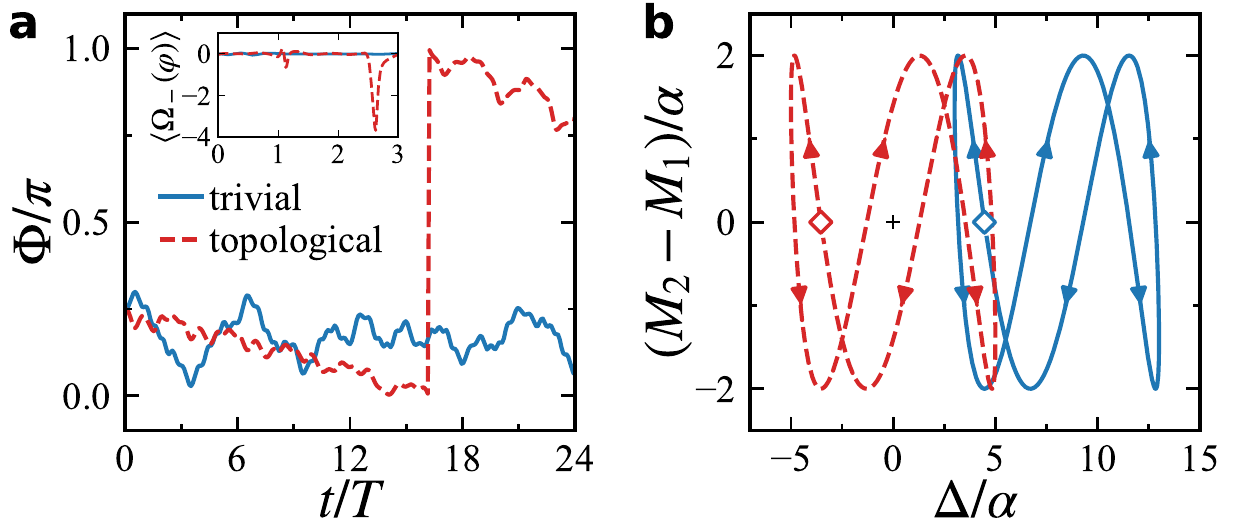}
	\caption{
		{\bf Dynamic evolution with the mismatched pump frequencies.}
		({\bf a}) Evolution of $\Phi$ for $\nu'=\nu/3$.
		The blue-solid (red-dashed) lines correspond to the trivial (topological) pump.
		The inner panel shows the evolution of $\langle\Omega_-(\varphi)\rangle$.
		The period of the pump cycling changes to $3T$ when $\nu'=\nu/3$.
		We set $\varphi_0=3\pi/4$ and $\Phi_i=\pi/4$.
		Other parameters are the same with Fig.\ref{fig:berry-curvature}.
		({\bf b}) The corresponding trajectories in the parameterized plane.
	}
	\label{fig:evol-mismatched-freq}
\end{figure}

If $\Delta$ and and $M_{1,2}$ in Eq.(\ref{eq:params-pump}) are generated by different experimental implements,
it is possible to individually tune the phase $\varphi_0$ as well as the pump frequencies $\nu$ and $\nu'$.
In Fig.\ref{fig:evol-mismatched-freq}, we study the case with mismatched $\nu$ and $\nu'$.
One can find the similar results as in the matched case.
The evolution trajectory for TMOE follows the Lissajous curve that encloses the singular point as shown in FIG.\ref{fig:evol-mismatched-freq}{\bf b}.
The net winding trajectory around the singular point is clockwise.
It reveals the sign of the topological MO rotation is negative, which is consistent with FIG.\ref{fig:evol-mismatched-freq}{\bf a}.

{~\\\noindent\bf \MakeUppercase{Discussions}}

The artificial TMOE can be realized using existing techniques of ultracold atoms.
For the alkali atoms like $^{87}$Rb,
the $N_s$-fold pseudo-spins can involve atomic ground states of both $|F=1\rangle\equiv|g\rangle$ and $|F=2\rangle\equiv|e\rangle$ levels \cite{Campbell2011pra,Wang2020npjqi}.
Specifically for the $N_s=4$ protocol shown in FIG.\ref{fig:setup}{\bf c},
we can choose two states from $|g\rangle$ as the pseudo-spin states $|2\rangle$ and $|4\rangle$,
and two from $|e\rangle$ as $|1\rangle$ and $|3\rangle$.
Since $|e\rangle$ usually suffers from strong losses,
the two coupling modes $M_1$ and $M_2$ between pseudo-spin states can be generated via the two-photon Raman process \cite{Yan2019prl},
which can support a time scale of millisecond order for experiments \cite{Blanshan2015pra}.
The relevant heating effect during the transition can be reduced under the far-detuning condition.
On the other hand, the alkaline-earth-metal(-like) atoms like $^{171}$Yb \cite{Safronova2018prl} have meta-stable excited states with long lifetime that exceeds milliseconds,
which can be an alternative candidate system for realizing TMOE.
In this case, the meta-stable states of alkaline-earth-metal atoms play the role of $|e\rangle$.
The energy offset $\Delta$ between $|g\rangle$ and $|e\rangle$ is naturally present via the detuning of the optical field coupling.
Alternatively, it can be introduced via the Stark shifts by using auxiliary fields.

The previous discussions focus on the noninteracting system.
In ultracold atoms,
the ubiquitous inter-atomic interaction invokes a nonlinear pump \cite{Jurgensen2021preprint}
and thus may affect the essential topological properties.
Practically, the interaction can be suppressed by far-off detuning the Feshbach resonance \cite{Kohler2006rmp,Chin2010rmp}.
On the other hand, the insight of TMOE is not limited to the topology of noninteracting systems.
Nontrivial attractive interaction, e.g. the $p$-wave form \cite{Zhang2008prl}, can give rise to topological superfluids \cite{Cooper2019rmp}.
If the emergent current is determined by the underlying topological invariant \cite{Hayward2018prb},
the atomic system can still be a candidate for realizing TMOE.

In summary, we have developed an experimentally feasible scheme for emulating and engineering novel TMOEs in ultracold atoms.
The artificial TMOE is signaled by the rotated polarization in the synthetic dimension, which is directly related to the bulk topology of the system. This proposal is very promising for exploring TMOEs in ultracold atoms.

{~\\\noindent\bf \MakeUppercase{Methods}}

Under the basis $|\psi\rangle = (\psi_1,\cdots,\psi_{N_s})^T$,
the evolution of the atoms is governed by the following Schr\"{o}dinger equation of Hamiltonian (\ref{eq:model-hamit}),
\begin{equation}
	i\partial_t |\psi\rangle = H |\psi\rangle \,.\label{eq:schrodinger-eq}
\end{equation}
Its numerical simulation is performed by using the QuTiP software package \cite{Johansson2012-qutip,Johansson2013-qutip}.

To continuously tune the initial polarized angle $\Phi_i$,
the atoms are loaded into two pseudo-spin states before entering the pumped area.
For simplicity without loss of generality, we choose the states $|1\rangle$ and $|2\rangle$ in the whole paper.
As $\hat{X}_{\rm cm}=n_1+in_2$ ($n_j$ denotes the density occupied in $|j\rangle$),
one can find that $\Phi_i$ ranges between 0 and $\pi/2$.
Specifically, $\Phi_i=\pi/4$ for the balanced occupation in the two pseudo-spin states.

{~\\\noindent\bf \MakeUppercase{Acknowledgments}}

We acknowledge S. Z. Zhang for helpful discussions.
This work was supported by the Key-Area Research and Development Program of Guangdong Province (Grant No. 2019B030330001), the NSFC / RGC JRS grant (No.: N\_HKU774/21), and the CRFs of Hong Kong (Nos. C6005-17G and C6009-20G).

{~\\\noindent\bf \MakeUppercase{Author contributions}}

Z.Z. performed the theoretical calculations.
Z.D.W. supervised the project.
All authors contributed to write the manuscript.

{~\\\noindent\bf \MakeUppercase{Competing interests}}

The authors declare no competing financial interests.

{~\\\noindent\bf \MakeUppercase{Data availability}}

The data that support this manuscript are available from the corresponding author upon reasonable request.

\vfill
%----------------------------------------------------------------------------------------

\end{document}